\documentclass[reprint,aps,prl,twocolumn]{revtex4-1}
\usepackage{amsmath}
\usepackage{amsfonts}
\usepackage{amssymb}
\usepackage{graphicx}
\usepackage[colorlinks,linkcolor=blue,anchorcolor=blue,citecolor=blue,urlcolor=black]{hyperref}
\usepackage{mathrsfs}
\usepackage{dcolumn}
\usepackage{bm}
\usepackage{epsfig}

\begin{document}

\title{Experimental demonstration of spontaneous chirality in a nonlinear microresonator}

\author{Qi-Tao Cao$^{1}$}
\author{He-Ming Wang$^{1}$}
\author{Chun-Hua Dong$^{2}$}
\author{Hui Jing$^{3}$}
\author{Rui-Shan Liu$^{1}$}
\author{Xi Chen$^{1}$}
\author{Li Ge$^{4,5}$}
\author{Qihuang Gong$^{1,6}$}
\author{Yun-Feng Xiao$^{1,6}$}

\affiliation{$^{1}$State Key Laboratory for Mesoscopic Physics and School of Physics, Peking University;
Collaborative Innovation Center of Quantum Matter, Beijing 100871, P. R. China}
\affiliation{$^{2}$Key Laboratory of Quantum Information, University of Science and Technology of China, Hefei 230026, P. R. China}
\affiliation{$^{3}$Department of Physics, Henan Normal University, Xinxiang 453007, P. R. China}
\affiliation{$^{4}$Department of Engineering Science and Physics, College of Staten Island, CUNY, Staten Island, New York 10314, USA}
\affiliation{$^5$The Graduate Center, CUNY, New York, NY 10016, USA}
\affiliation{$^{6}$Collaborative Innovation Center of Extreme Optics, Taiyuan 030006, Shanxi, P. R. China}

\date{\today}

\maketitle

\textbf{Chirality is an important concept that describes the asymmetry property of a system, 
which usually emerges spontaneously due to mirror symmetry breaking \cite{SSB.Malomed.book,SSB.Heil.PRL.2001,SSB.Green.PRL.1990}.
Such spontaneous chirality manifests predominantly as parity breaking in modern physics, which has been studied extensively, for instance, in Higgs physics \cite{C.Endres.Nat.2012}, double-well Bose-Einstein condensates \cite{C.Zibold.PRL.2010,C.Abbarchi.NPhys.2013}, topological insulators and superconductors \cite{C.Matano.NPhys.2016}.
In the optical domain, spontaneous chiral symmetry breaking has been elusive experimentally, especially for micro- and nano-photonics which demands multiple identical subsystems, such as photonic nanocavities \cite{C.Hamel.NPhot.2015}, meta-molecules \cite{C.Liu.NComm.2015} and other dual-core settings \cite{C.Kevrekidis.PLA.2005,C.Snyder.JOSAB.1991}.
Here, for the first time, we observe spontaneous emergence of a chiral field in a single ultrahigh-$Q$ whispering-gallery microresonator. 
This counter-intuitive effect arises due to the inherent Kerr nonlinearity-modulated coupling between clockwise (CW) and counterclockwise (CCW) propagating waves.
At an ultra-weak input threshold of a few hundred microwatts, the initial chiral symmetry is broken spontaneously, and the CW-to-CCW output ratio reaches 20:1.
The spontaneous chirality in microresonator holds great potential in studies of fundamental physics and applied photonic devices.}

\begin{figure}
\includegraphics[width=85mm,clip]{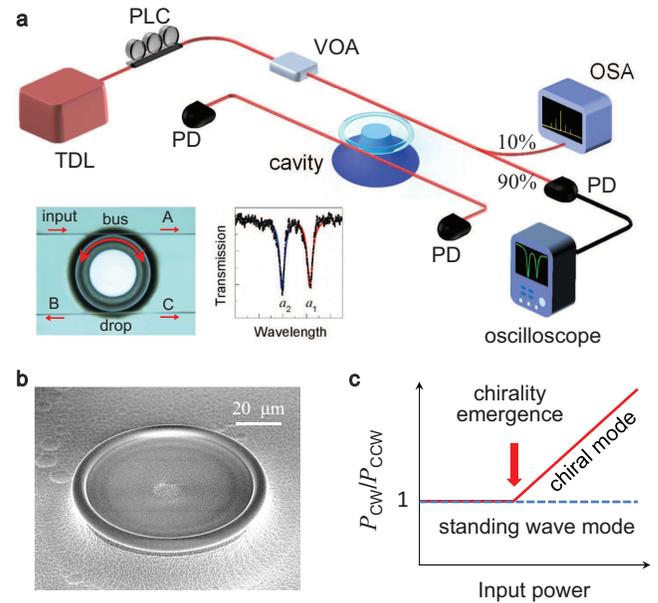}
\caption{\label{cavity}
\textbf{Experimental setup.}
\textbf{a}, A toroidal microcavity is evanescently side coupled by two tapered fibre waveguides. Left inset: the bus waveguide is used to excite WGMs, and ports B and C of the drop waveguide collect the emission of CW and CCW components separately. Right inset: a typical transmission spectrum collected by port A.
TDL: tunable diode laser; PLC: polarization controller; VOA: variable optical attenuator; 
OSA: optical spectroscope analyzer; PD: photodetector.
\textbf{b}, Scanning electron microscope image of the silica microtoroid.
\textbf{c}, Schematic illustration of CW to CCW intensity ratio for a standing wave mode (blue) and a CW chiral mode (red), as the input power increases.
}
\end{figure}

\begin{figure*}
\includegraphics[width=130mm]{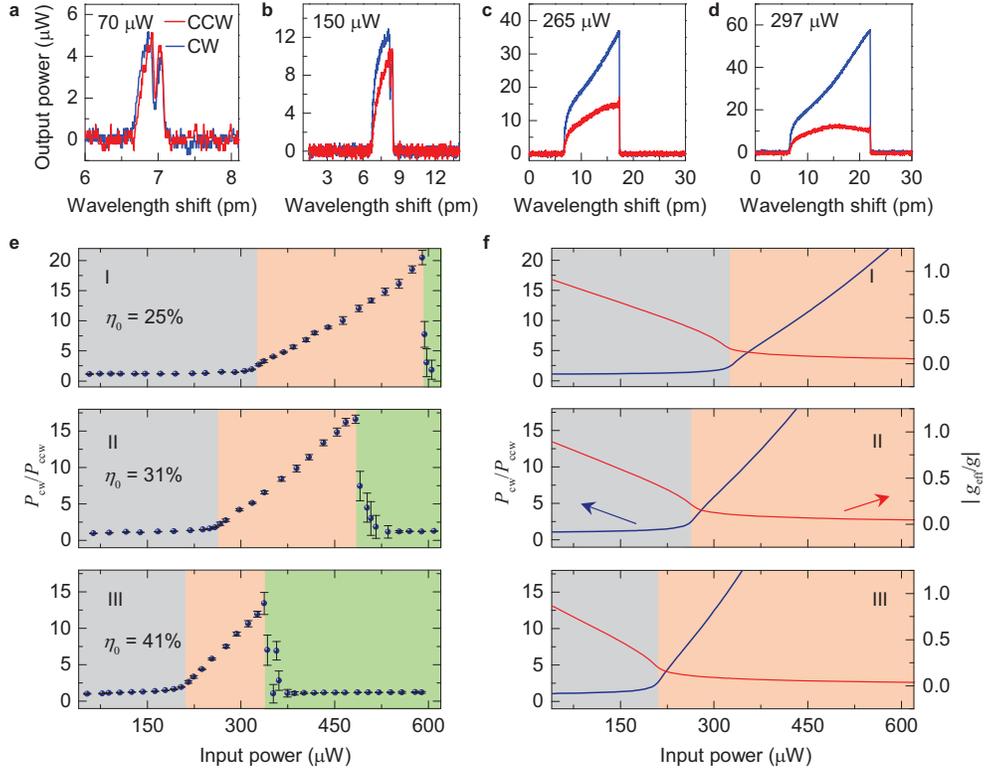}
\caption{\label{ratio}
\textbf{Onset of the chirality with increasing input power.}
\textbf{a-d}, The CW (blue) an CCW (red) output spectra collected by the drop fibre under the inputs of $70$ $\mathrm{\mu W}$, $150$ $\mathrm{\mu W}$, $265$ $\mathrm{\mu W}$ and $297$ $\mathrm{\mu W}$.
\textbf{e}, The experimental ratio $\mathfrak{R} = P_\mathrm{cw}/P_\mathrm{ccw}$ versus the input power, with the initial coupling efficiency of $\eta_0=25\%$, $31\%$ and $41\%$, from top to bottom.
\textbf{f}, The theoretical ratio $\mathfrak{R}$ (blue curve) and the scaled effective coupling strength $|g_\mathrm{eff}/g|$ (red curve) under the same experimental condition in \textbf{e}.
}
\end{figure*}

The ultrahigh-$Q$ whispering-gallery mode (WGM) microresonator is an indispensable optical system for applications in a wide range of fields ranging from strong-coupling cavity quantum electrodynamics, cavity optomechanics, ultralow-threshold lasers to highly sensitive sensing \cite{WGM.Cao.RMP.2015}. In a practical WGM resonator, the CW and CCW propagating waves are coupled to each other due to the backscattering, leading to symmetric and antisymmetric standing waves with equal CW and CCW amplitudes \cite{WGM.Weiss.OL.1995,WGM.Gorodetsky.JOSAB.2000,WGM.Kippenberg.OL.2002,WGM.Mazzei.PRL.2007,WGM.Zhu.NPhot.2010}.
While this mirror symmetry can be broken locally by wave effects such as the Goos-H\"anshen shift and Fresnel filtering for certain regions of an optical mode
\cite{SymP.Song.PRL.2010,SymP.Redding.PRL.2012}, the demonstrations of overall optical chirality have to rely on external perturbations to a single resonator, 
either by breaking the mirror \cite{SymP.Chern.APL.2003,SymP.Song.SRep.2014,SymP.Wiersig.PRL.2014,SymP.Sarma.PRL.2015,SymP.Peng.PNAS.2016} or time-reversal \cite{SymT.Wang.Nature.2009,SymT.Ge.Optica.2015,SymT.Kim.NPhys.2015} symmetry.
Such chirality with unbalanced CW and CCW components not only attracts general interest in physics, but also is of importance in novel devices such as unidirectional-emission microlasers \cite{SymP.Chern.APL.2003,SymP.Song.PRL.2010,SymP.Redding.PRL.2012,SymP.Song.SRep.2014}, optical gyroscopes \cite{SymP.Sarma.PRL.2015,SymT.Ge.Optica.2015,SymT.Kim.NPhys.2015}, and single-particle detection \cite{SymP.Wiersig.PRL.2014,SymP.Peng.PNAS.2016}.
In this Letter, we experimentally demonstrate the spontaneous emergence of a chiral optical field in a single ultrahigh-Q WGM microresonator (Fig. \ref{cavity}) without any breaking of spatial- or time-reversal symmetry.
Above a threshold input, the initial balanced CW and CCW components of WGM field become highly unbalanced as schematically illustrated in Fig. \ref{cavity}c.

As shown in Fig. \ref{cavity}a, b, a silica toroidal microresonator with a principal (minor) diameter of $78$ $\mathrm{\mu m}$ ($6.8$ $\mathrm{\mu m}$) is side coupled by a bus and a drop fibre waveguides.
In the 1550 nm wavelength band, two nearby resonances with ultra-high quality factor ($Q\sim7.8$$\times$10$^{7}$), a low-frequency symmetric mode $a_1$ and a high-frequency antisymmetric mode $a_2$, can be seen in the transmission spectrum (inset of Fig. \ref{cavity}a). 
Under the input from the bus fibre, the intensities of CW and CCW components are measured by the drop fibre. It is clear that each mode of this doublet is chiral symmetric when the input power is weak, containing equal amounts of CW and CCW components (Fig. \ref{ratio}a).

\begin{figure*}
\includegraphics[width=120mm]{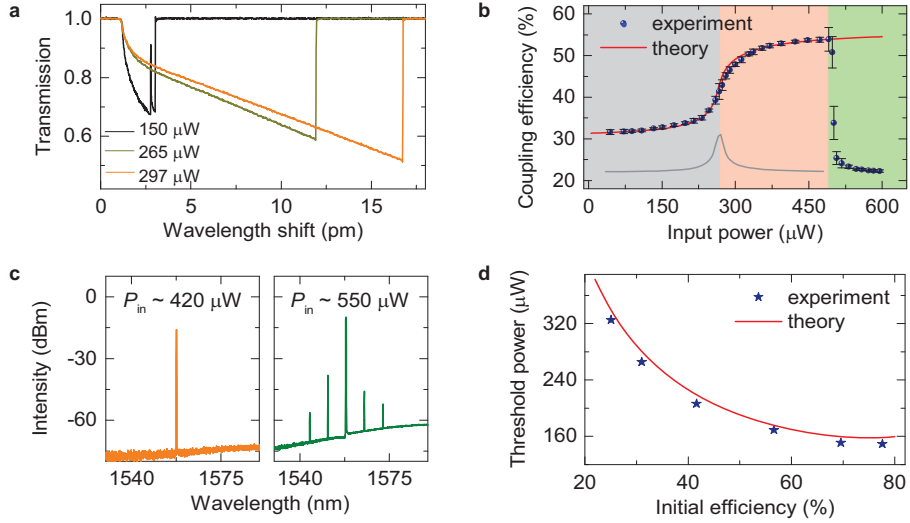}
\caption{\label{efficiency}
\textbf{Coupling efficiency and spontaneous chirality.}
\textbf{a}, Typical transmission spectra for the input power of $150$ $\mathrm{\mu W}$, $265$ $\mathrm{\mu W}$, and $297$ $\mathrm{\mu W}$, at a fixed coupling position. The on-resonance coupling efficiencies $\eta$ between the bus waveguide and the cavity rise from $32 \%$ to $50\%$ as the input increases.
\textbf{b}, The coupling efficiency $\eta$ versus the input power. The blue dots are experimental data, and the red curve is the theoretical result. The grey curve (vertical axis not shown here) plots the derivative of the empirical fitting from experimental data, the maximum of which corresponds to the threshold power.
\textbf{c}, Optical spectra under the input of $420$ $\mathrm{\mu W}$ (left) and $550$ $\mathrm{\mu W}$ (right).
\textbf{d}, Dependence of experimental (blue stars) and analytical (red curve) thresholds on the initial coupling efficiency $\eta_0$.
}
\end{figure*}

As we increase the input power, chiral symmetry breaking occurs spontaneously. Although the spectra of the intensities collected by the drop fibre no longer have a Lorentzian shape, the growing imbalance between the CW and CCW components are readily seen in Figs. \ref{ratio}b-d.
To be quantitative, we characterize the chirality by the ratio $\mathfrak{R} = P_\mathrm{cw}/P_\mathrm{ccw}$, where the $P_\mathrm{cw}$ ($P_\mathrm{ccw}$) is the on-resonance output power of the CW (CCW) component at the peak of the broadened spectrum. It is evident that $\mathfrak{R}$ shows a threshold behavior as the input power increases (see Fig. \ref{ratio}e-II for example): $\mathfrak{R}$ is near unity with the input power from $40$ $\mathrm{\mu W}$ to $260$ $\mathrm{\mu W}$, and it increases rapidly after $260$ $\mathrm{\mu W}$. We note that the suddenly drop of $\mathfrak{R}$ near $500$ $\mathrm{\mu W}$ is due to four-wave mixing (FWM), as we show in Fig.~\ref{efficiency}c.

This spontaneous chiral symmetry breaking is found to be a universal behavior in our system. More specifically, it is independent of the initial coupling efficiency $\eta_0$ between the bus fibre and the microresonator at weak input power. The $\eta_0$ varies from sample to sample, but the threshold behavior of $\mathfrak{R}$ persists as we show in Figs. \ref{ratio}e-I to III. We further note that the threshold power increases as the bus fibre-cavity coupling decreases (see also Fig.~\ref{efficiency}d), so does the power range of chiral modes before the emergence of FWM (see Supplementary Information). The latter leads to a higher maximal value of $\mathfrak{R}$ at a lower $\eta_0$, which reaches about 20 when $\eta_0=25\%$.

Remarkably, the threshold power for spontaneous chiral symmetry breaking also manifests itself in the power dependence of the fibre coupling efficiency $\eta$, which can be obtained by analyzing the transmission spectra. The latter shows the same broadening and distortion with increased power as the intensity spectra collected in the bus fibre (Fig. \ref{efficiency}a). These changes are expected by considering the thermal effect and Kerr nonlinearity \cite{Th.Carmon.OE.2004}.
Note that although the thermal effect causes the red-shift and mode broadening, it does not affect the coupling efficiency (see Supplementary Information).
Differing from the previous studies, the on-resonance coupling becomes stronger.
The dependence of the coupling efficiency $\eta$ on the input power is plotted in Fig. \ref{efficiency}b (blue dots).
It displays an $S$ curve before the FWM regime, and its inflection point (where its slope reaches the maximum, shown by the grey curve in Fig. \ref{efficiency}b) is exactly at the chiral threshold.

\begin{figure*}
\includegraphics[width=125mm]{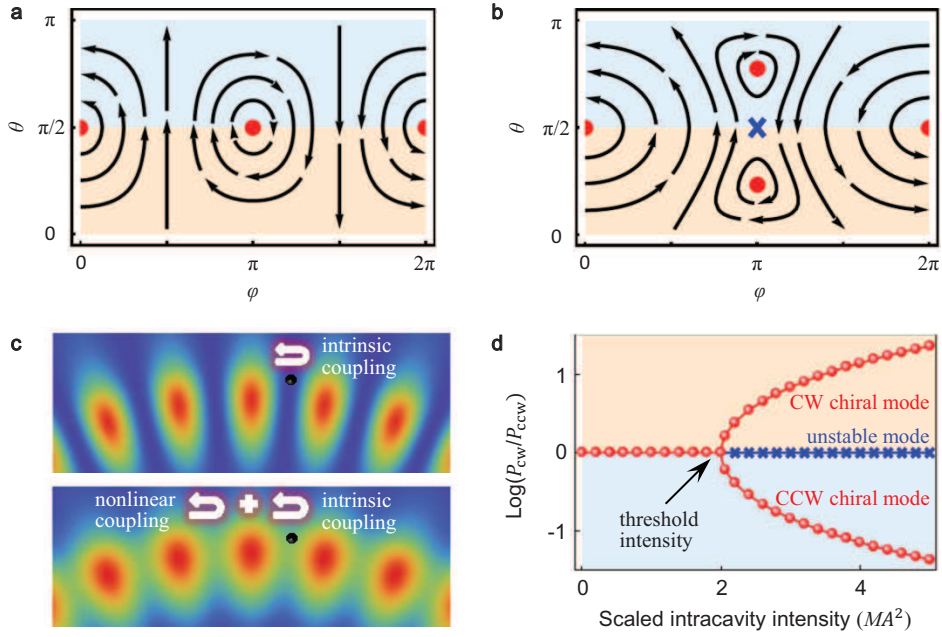}
\caption{\label{PS1}
\textbf{Phase diagram and bifurcation.}
\textbf{a}, \textbf{b}, Trajectories of state evolutions for $MA^2=0$, $3$ in the phase space.
States dominated by the CCW (CW) wave components lie in the upper blue-shaded area (lower orange-shaded area), with $\pi/2<\theta\leq\pi$ ($0\leq\theta<\pi/2$).
\textbf{c}, Patterns of the standing wave mode (top) and the chiral mode (bottom) (see Supplementary Information).
In the former, a nanoparticle (black dot) produces the intrinsic coupling between CW and CCW components,
while in the latter the nonlinear coupling is present.
\textbf{d}, The ratio $\mathfrak{R} =P_\mathrm{cw}/P_\mathrm{ccw}$  depending on the the scaled intracavity intensity $MA^2$.
In \textbf{a}, \textbf{b}, and \textbf{d}, red dots (blue crosses) mark stable (unstable) modes.
}
\end{figure*}

We now analyze the mode dynamics of a lossless cavity by introducing Kerr nonlinearity. The Hamiltonian of the cavity reads (see Supplementary Information)
$H=\hbar g(a_2^\dag a_2-a_1^\dag a_1)-\frac{1}{4}(M\hbar \omega) \hbar g (\delta_{\mu\nu}\delta_{\rho\sigma}+\delta_{\mu\sigma}\delta_{\nu\rho}+\delta_{\mu\rho}\delta_{\nu\sigma})a_\mu^\dag a_\nu a_\rho^\dag a_\sigma$,
where 
$g$ is the coupling strength between CW and CCW waves,
$\omega$ is the mode frequency,
$\delta$ is the Kronecker delta,
and repeated indices are summed over.
The coefficient $M$ is proportional to the scalar Kerr nonlinear susceptibility $\chi^{(3)}$.
The state amplitudes can be rewritten in the CW-CCW basis with four real parameters, $a_{\mathrm{cw}}=(a_1+i a_2)/\sqrt{2}\equiv Ae^{i\alpha}\cos(\theta/2)$, $a_{\mathrm{ccw}}=(a_1-i a_2)/\sqrt{2}\equiv Ae^{i\alpha}\sin(\theta/2)e^{i\varphi}$,
where $Ae^{i\alpha}$ is the total complex amplitude,
and the coupled-mode equations read
\begin{equation}
\label{geff}
\frac{1}{ig}\frac{da_{m}}{dt}=MA^2a_{m}+(1+Ma_{m'}^*a_{m})a_{m'},
\end{equation}
where $m$ and $m'$ ($m\neq m'$) stand for CW and CCW.
The Bloch sphere parameters, $0\leq\theta\leq\pi$ and $0\leq\varphi\leq 2\pi$, describe the relative amplitude and phase difference of the two propagating waves, which can be used to study the evolution of states and existing modes. As shown in Fig. \ref{PS1}, 
the CW and CCW amplitudes of the states oscillate because of the intrinsic coupling.
In the case of $MA^2=0$ where the nonlinear effects are turned off (Fig. \ref{PS1}a), the trajectories circulate around the red dots at $\theta=\pi/2$, $\varphi=0$ and $\theta=\pi/2$, $\varphi=\pi$.
The former fixed point corresponds exactly to the symmetric mode $a_1$ while the latter to the antisymmetric mode $a_2$.
After the intensity increases to $MA^2=2$, a pitchfork bifurcation occurs and two new modes emerge (e.g., Fig. \ref{PS1}b at $MA^2=3$).
These new modes (marked as two red dots at $\varphi=\pi$) gain chirality spontaneously, 
while the mode $a_2$ at $\theta=\pi/2$, $\varphi=\pi$ (marked as a blue cross) becomes unstable.

As suggested by Eq. (\ref{geff}), the coupling between the CW and CCW waves consists of both the intrinsic coupling and the nonlinear intermodal interaction (Fig. \ref{PS1}c) resulting in an {\it intensity-dependent} effective coupling coefficient $g_\mathrm{eff}\equiv(1+Ma_\mathrm{ccw}^*a_\mathrm{cw})g$.
Considering the optical Kerr effect, the standing-wave modes can be considered as periodic potentials along the microcavity perimeter.
In this regard, the propagating waves are reflected by the nonlinearity-related potential, adding coherently to the intrinsic coupling.
With a low intracavity intensity, the nonlinear modulation on the refractive index is weak, and $g_\mathrm{eff}\sim g>0$.
Thus, the unbalanced CW and CCW components oscillate for a chiral state, and cannot become stable to form a mode.
With a strong enough intensity above the threshold $A^2=2M^{-1}$, the coupling between the propagating waves is canceled effectively, i.e., $g_\mathrm{eff}=0$, under the phase-matching condition $\varphi=\pi$ satisfied by the antisymmetric mode.
In this case, the chiral symmetry is broken.
Figure \ref{PS1}d describes the ratio $\mathfrak{R} =P_\mathrm{cw}/P_\mathrm{ccw}$ depending on the intracavity intensity $A^2$, under the condition $\varphi=\pi$.
When the intracavity intensity exceeds the threshold value, bifurcation occurs and chiral modes emerge. The CW (CCW) chiral mode can be found on the upper (lower) branch with Log$\mathfrak{R}>0$ (Log$\mathfrak{R}<0$), while the original antisymmetric mode becomes unstable.
A stronger intensity leads to a greater chirality.
For the low-frequency mode $a_1$ corresponding to $\varphi=0$, however, the coupling is enhanced by the nonlinear effects, i.e., $g_\mathrm{eff}>g$, and no chirality can be found near this mode.

In a real system, both loss and input are present, and the corresponding intensity ratio $\mathfrak{R}$ and coupling efficiency $\eta$ can be derived (see Supplementary Information) as plotted in Figs. \ref{ratio}f and \ref{efficiency}b.
It is found that the experimental results are in good agreement with the theory until FWM appears.
For instance, at the threshold value of $260$ $\mathrm{\mu W}$ in Figs. \ref{ratio}e-II and \ref{efficiency}b, the bifurcation occurs and the chiral symmetry is broken spontaneously, where $|g_\mathrm{eff}|$ decreases to zero.
With increasing input power, the system enters a highly chiral state with ratio $\mathfrak{R}$ of $~20$ in our experiment.
Here the CW chiral mode is excited because the phase of its electric field matches that of the input much better.

The Kerr nonlinearity-induced spontaneous chirality in a single microresonator, without any explicit breaking of spatial- or time-reversal symmetry,
is different from the unidirectional ring laser which relies on the carrier concentration and the mode competition \cite{RL.Sorel.IEEE.2004,RL.Zhukovsky.PRA.2009}.
We also note that our chirality threshold is not at an exceptional point \cite{EP1}, which would otherwise introduce a square root singularity to the eigenvalues of the effective Hamiltonian represented by Eq.~(1). The latter, however, is not the case as we show in the Supplementary Information, hence the pitchfork bifurcation at the chirality threshold, though seemingly identical to parity-time symmetry breaking \cite{PT.Bender.PRL.1998}, is caused by a very different mechanism.
Furthermore, although the chirality vanishes once the input power is strong enough to generate side bands via FWM, the latter can be suppressed by delicate designs of the cavity geometry.

\noindent\textbf{METHODS.}

\noindent\textbf{Device fabrication.}
The circular microtoroid is fabricated from a 2 $\mathrm{\mu m}$-thick layer of silica on a silicon wafer \cite{fab.Armani.Nat.2003}.
A pair of parallel tapered fibres are prepared using a hydrogen flame, which are separated by over $150$ $\mathrm{\mu m}$.
A 3-axis nano-translation stage (Thorlabs, MDT630A) accurately controls the coupling between the bus waveguide and the resonator.
After that, a piece of clean silica-chip touching the drop fibre waveguide is used to control the position of this fibre, and we can adjust the silica-chip in two dimensions to control the coupling between the drop waveguide and the resonator.

\noindent\textbf{Experimental implementation.}
The input light is from a tunable diode laser (New Focus, TLB 6328, 1520-1570 $\mathrm{nm}$).
A triangular wave voltage is applied on the piezoelectric transducer of the diode lasers to sweep the wavelength of the input light with the speed of 36.4 $\mathrm{nm/s}$.
The input power can be adjusted by controlling the voltage of the MEMS VOA (Agiltronn, TMOA-115211332),
which does not affect the polarization of the input light.
We sweep the input power from 40 $\mathrm{\mu W}$ to 650 $\mathrm{\mu W}$ by applying the triangular wave voltage on the VOA with a cycle of 200 $\mathrm{s}$.
Three 125 $\mathrm{MHz}$ photodetectors are used for the collections of the cavity emissions from ports A, B, C.
The experimental data are recorded on a DAQ board (National Instruments, PCI-6115) with a 10 $\mathrm{MHz}$ sampling frequency, and an OSA (YOKOGAWA, AQ6319, 600-1700 $\mathrm{nm}$) is used to show the optical spectra with a resolution of 0.05 $\mathrm{nm}$ simultaneously.

\addvspace{0.5cm}
\noindent \textbf{Acknowledgments}

\noindent This project was supported by the 973 program (Grant Nos. 2013CB921904 and 2013CB328704) and the NSFC (Grant Nos. 61435001, 11474011, and 11222440).
H.-M. W., X. C. and R.-S. L. were supported by the National Fund for Fostering Talents of Basic Science (Grants Nos. J1030310 and J1103205).

\addvspace{0.5cm}
\noindent \textbf{Author contributions}

\noindent Q.T.C., H.M.W. and C.H.D. contributed equally. Q.T.C. and C.H.D. performed the experiment. H.M.W. built the theoretical model. Y.F.X. designed the experiment and supervised the project. All authors contributed to the discussion, analyzed the data and wrote the manuscript. Correspondence and requests for materials should be addressed to Y.F.X. (Email: yfxiao@pku.edu.cn)


\begin{thebibliography}{99}

\bibitem{SSB.Malomed.book} Malomed, B. A.
Spontaneous Symmetry Breaking, Self-Trapping, and Josephson Oscillations.
(Springer, 2013).

\bibitem{SSB.Heil.PRL.2001} Heil, T., Fischer, I., Elsa\"sser, W., Mulet, J. \& Mirasso, C. R.
Chaos synchronization and spontaneous symmetry-breaking in symmetrically delay-coupled semiconductor lasers.
{\it Phys. Rev. Lett.} {\bf 86}, 795 (2001).

\bibitem{SSB.Green.PRL.1990} Green, C., Mindlin, G. B., D'Angelo, E. J., Solari, H. G. \& Tredicce, J. R
 Spontaneous symmetry breaking in a laser: the experimental side.
{\it Phys. Rev. Lett.} {\bf 65}, 3124 (1990).

\bibitem{C.Endres.Nat.2012} Endres, M. {\it et al.}
The `Higgs' amplitude mode at the two-dimensional superfluid/Mott insulator transition.
{\it Nature} {\bf 487}, 454--458 (2012).

\bibitem{C.Zibold.PRL.2010} Zibold, T., Nicklas, E., Gross, C. \& Oberthaler, M. K.
Classical bifurcation at the transition from Rabi to Josephson dynamics.
{\it Phys. Rev. Lett.} {\bf 105}, 204101 (2010).

\bibitem{C.Abbarchi.NPhys.2013} Abbarchi, M. \textit{et al.}
Macroscopic quantum self-trapping and Josephson oscillations of exciton polaritons.
{\it Nature Phys.} {\bf 9}, 275--279 (2013).

\bibitem{C.Matano.NPhys.2016} Matano, K., Kriener, M., Segawa, K., Ando, Y. \& Zheng, G. Q.
Spin-rotation symmetry breaking in the superconducting state of $\mathrm{Cu_{x}Bi_{2}Se_{3}}$.
{\it Nature Phys.} doi: 10.1038/nphys3781 (2016).

\bibitem{C.Hamel.NPhot.2015} Hamel, P. {\it et al.}
Spontaneous mirror-symmetry breaking in coupled photonic-crystal nanolasers.
{\it Nature Photon.} {\bf 9}, 311--315 (2015).

\bibitem{C.Liu.NComm.2015} Liu, M., Powell, D. A., Shadrivov, I. V., Lapine, M. \& Kivshar, Y. S. Spontaneous chiral symmetry breaking in metamaterials. {\it Nature Commun.} {\bf 5}, 4441 (2014).

\bibitem{C.Kevrekidis.PLA.2005} Kevrekidis, P. G., Chen, Z., Malomed, B. A., Frantzeskakis, D. J. \& Weinstein, M. I.
Spontaneous symmetry breaking in photonic lattices: theory and experiment.
{\it Phys. Lett. A} {\bf 340}, 275--280 (2005).

\bibitem{C.Snyder.JOSAB.1991} Snyder, A.W., Mitchell, D.J., Poladian, L., Rowland, D.R. \& Chen, Y.
Physics of nonlinear fiber couplers.
{\it J. Opt. Soc. Am. B} {\bf 8}, 2101--2118 (1991).

\bibitem{WGM.Cao.RMP.2015} Cao, H. \& Wiersig, J.
Dielectric microcavities: Model systems for wave chaos and non-Hermitian physics.
{\it Rev. Mod. Phys.} {\bf 87}, 61 (2015).

\bibitem{WGM.Weiss.OL.1995} Weiss, D. S. {\it et al.}
Splitting of high-Q Mie modes induced by light backscattering in silica microspheres.
{\it Opt. Lett.} {\bf 20}, 1835--1837 (1995).

\bibitem{WGM.Gorodetsky.JOSAB.2000} Gorodetsky, M. L., Pryamikov, A. D. \& Ilchenko, V. S.
Rayleigh scattering in high-$Q$ microspheres.
{\it J. Opt. Soc. Am. B} {\bf 17}, 1051--1057 (2000).

\bibitem{WGM.Kippenberg.OL.2002} Kippenberg, T. J., Spillane, S. M. \& Vahala, K. J.
Modal coupling in traveling-wave resonators.
{\it Opt. Lett.} {\bf 27}, 1669--1671 (2002).

\bibitem{WGM.Mazzei.PRL.2007} Mazzei, A. {\it et al.}
Controlled Coupling of Counterpropagating Whispering-Gallery Modes by a Single Rayleigh Scatterer: A Classical Problem in a Quantum Optical Light.
{\it Phys. Rev. Lett.} {\bf 99}, 173603 (2007).

\bibitem{WGM.Zhu.NPhot.2010} Zhu, J. {\it et al.}
On-chip single nanoparticle detection and sizing by mode splitting in an ultrahigh-$Q$ microresonator.
{\it Nature Photon.} {\bf 4}, 46--49 (2010).




\bibitem{SymP.Song.PRL.2010} Song, Q. H. {\it et al.}
Directional Laser Emission from a Wavelength-Scale Chaotic Microcavity.
{\it Phys. Rev. Lett.} {\bf 105}, 103902 (2010).

\bibitem{SymP.Redding.PRL.2012} Redding, B. {\it et al.}
Local Chirality of Optical Resonances in Ultrasmall Resonators.
{\it Phys. Rev. Lett.} {\bf 108}, 253902 (2012).

\bibitem{SymP.Chern.APL.2003} Chern, G. D., Tureci, H. E., Stone, A. D., Chang, R. K., Kneissl, M. \& Johnson, N. M.
Unidirectional lasing from InGaN multiple-quantum-well spiral-shaped micropillars.
{\it Appl. Phys. Lett.} {\bf 83}, 1710--1712 (2003).

\bibitem{SymP.Song.SRep.2014} Song, Q. {\it et al.}
The combination of high Q factor and chirality in twin cavities and microcavity chain.
{\it Sci. Rep.} {\bf 4}, 6493 (2014).

\bibitem{SymP.Sarma.PRL.2015} Sarma, R., Ge, L., Wiersig, J. \& Cao, H.
Rotating Optical Microcavities with Broken Chiral Symmetry.
{\it Phys. Rev. Lett.} {\bf 114}, 053903 (2015).

\bibitem{SymP.Wiersig.PRL.2014} Wiersig, J.
Enhancing the Sensitivity of Frequency and Energy Splitting Detection by Using Exceptional Points: Application to Microcavity Sensors for Single-Particle Detection.
{\it Phys. Rev. Lett.} {\bf 112}, 203901 (2014).


\bibitem{SymP.Peng.PNAS.2016} Peng, B. \textit{et al}. Chiral modes and directional lasing at exceptional points. \textit{PNAS}, in press (2016).

\bibitem{SymT.Ge.Optica.2015} Ge, L., Sarma, R., \& Cao, H.
Rotation-induced Evolution of Far-field Emission Patterns of Deformed Microdisk Cavities.
{\it Optica} {\bf 2}, 323--328 (2015).

\bibitem{SymT.Kim.NPhys.2015} Kim, J., Kuzyk, M. C., Han, K., Wang, H. \& Bahl G.
Non-reciprocal Brillouin scattering induced transparency.
{\it Nature Phys.} {\bf 11}, 275--280 (2015).

\bibitem{SymT.Wang.Nature.2009} Wang, Z., Chong, Y., Joannopoulos, J. D. \& Solja\v{c}i\'{c}, M. Observation of unidirectional backscattering-immune topological electromagnetic states. \textit{Nature} \textbf{461}, 772--775 (2009).



\bibitem{Th.Carmon.OE.2004} Carmon, T., Yang, L. \& Vahala, K. J.
Dynamical thermal behavior and thermal selfstablity of microcavities.
{\it Opt. Express} {\bf 12}, 4742--4750 (2004).

\bibitem{RL.Sorel.IEEE.2004} Sorel, M. {\it et al.}
Operating regimes of GaAs-AlGaAs semiconductor ring lasers: experiment and model.
{\it IEEE J. Quantum Electron.} {\bf 39}, 1187--1195 (2004).

\bibitem{RL.Zhukovsky.PRA.2009} Zhukovsky, S. V., Chigrin, D. N., \& Kroha, J.
Bistability and mode interaction in microlasers.
{\it Phys. Rev. A.} {\bf 79} (2009).



\bibitem{EP1} Okolowicz, J. Ploszajczak, M. \& Rotter, I.
Dynamics of quantum systems embedded in a continuum.
\textit{Phys. Rep.} {\bf 374}, 271 (2003).

\bibitem{PT.Bender.PRL.1998} Bender, C. M., \& Boettcher, S.
Real spectra in non-Hermitian Hamiltonians having $\mathcal{P T}$ symmetry.
{\it Phys. Rev. Lett.} {\bf 80}, 5243 (1998).


\bibitem{fab.Armani.Nat.2003} Armani, D. K., Kippenberg, T. J., Spillane, S. M. \& Vahala, K. J.
Ultra-high-Q toroid microcavity on a chip.
{\it Nature} {\bf 421}, 925--928 (2003).

\end{thebibliography}
\end{document}